\documentclass[aps,pra,reprint,groupedaddress,showpacs]{revtex4-1}
\usepackage{graphicx}% Include figure files
\usepackage{dcolumn}% Align table columns on decimal point
\usepackage{bm}% bold math
\usepackage{amssymb}
\usepackage{mathrsfs}

\begin{document}
%\begin{CJK}{GBK}{song}
%\preprint{}

\title{Theoretical study of the quantum noise in phase-sensitive heterodyne detection with a bichromatic local oscillator}


\author{Sheng Feng}
\email[]{fengsf2a@hust.edu.cn}
\author{Dechao He}
\author{Heng Fan}
%\author{Yu Xiao}
%\thanks{}
\affiliation{
MOE Key Laboratory of Fundamental Quantities Measurement, School of Physics, Huazhong University of Science and Technology, Wuhan 430074, China}


\date{\today}

\begin{abstract}
A traditional heterodyne detector, as a phase-insensitive device, suffers the well-known 3 dB noise penalty caused by image sideband vacuum. In contrast, a heterodyne detector with a bichromatic local oscillator, as a phase-sensitive device, should be exempted from the 3 dB noise penalty, in spite of the existence of the image sideband vacuum. Assuming coherent light at the input, we develop in this work a theory to describe the quantum nature of the phase-sensitive heterodyne detector, in a good agreement with experiment. The absence of the quantum noise of the image vacuum modes in the heterodyne detector may be explained by that the studied detector senses only a single field of light, i.e., the signal field, according to the theory developed.
% are widely-used tools to sense weak optical signals in the context of classical optics. However, the well-known 3 dB heterodyne quantum noise, introduced by the image sideband vacuum, makes them less competitive than their homodyne counterparts in quantum optics.
%confirms that the studied heterodyne detector has the potential of being a useful tool for precision measurements with squeezed light. In addition, this work is important for us to understand the origin for the quantum noise in heterodyne detection.
%In a recently accomplished experiment, we discovered that the 3 dB heterodyne noise was absent, or significantly suppressed, in phase-sensitive heterodyne detection with a bichromatic local oscillator. As part of the effort to understand the mechanism for the absence, or suppression, of the heterodyne noise, 
%The quantum theory of optical detection utilizes the essential concept of image sideband mode to describe the quantum behavior of heterodyne detectors, which have obvious advantages over homodyne ones for precision measurements in the context of classical optics. It remained an experimental challenge to suppress the 
%in optical heterodyning, until recently when we experimentally discovered that the 3 dB extra quantum noise was significantly suppressed 
% Otherwise, a theoretical dilemma would come into being in the quantum theory of optical detection. 
\end{abstract}

\pacs{42.50.Lc, 42.50.Xa 42.50.Dv}
%\keywords{}

\maketitle

\section{Introduction}
The concept of image sideband vacuum mode plays a crucial role in our orthodox understanding of the quantum nature of heterodyne detectors \cite{personick1971,yuen1978,shapiro1979,yuen1980,yuen1983,schumaker1984,yamamoto1986,collett1987,carmichael1987,ou1987,caves1994,ou1995,haus2011}. A traditional heterodyne detector, as a phase-insensitive device, suffers a 3 dB noise penalty caused by the extra quantum noise of the image sideband vacuum mode involved in the detection \cite{personick1971,yuen1980,yuen1983,yamamoto1986,caves1994}. This 3 dB heterodyne noise makes heterodyne detectors less competitive than their homodyne counterparts in the application of precision measurements with non-classical light.

A proposal has been put forth to suppress the 3 dB heterodyne quantum noise by replacing the image sideband vacuum mode with light in two-photon coherent states \cite{yuen1980}. However, this proposal has never been experimentally realized. Another possible solution to the problem is to excite the image sideband mode into a coherent state at a power level similar to the signal mode. Thereby, the heterodyne detector becomes a phase-sensitive one and is free of the 3 dB noise penalty \cite{collett1987,caves1982}. The shortcoming of the second method is that the input signal must be pre-processed before it is received by the detector. Moreover, the gained signal-to-noise ratio (SNR) results from doubling the power of the signal before detection and the quantum noise floor of the heterodyne detector remains the same.

\begin{figure} 
\includegraphics[scale=0.35]{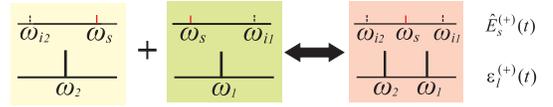}
\caption{\label{fig:pih2psh} (color online) Illustration of the idea to construct a phase-sensitive heterodyne detector out of two traditional phase-insensitive heterodyne ones. The idea can be realized in practice by utilizing a bichromatic local oscillator, whose only nonzero components are two coherent modes of frequency at $\omega_1$ and $\omega_2$, for the detector. Obviously, the two image sideband modes involved in the detection at frequencies $\omega_{i1}$ and $\omega_{i2}$ are in vacuum states. The center-frequency mode in the phase-sensitive detection, labeled by red color, is the signal mode at frequency $\omega_s$. $\hat{E}_s^{(+)}$ and $\mathscr{E}_l^{(+)}(t)$ stand for the signal field and the oscillator field, respectively.
}
\end{figure}

We study a phase-sensitive heterodyne detector with a bichromatic local oscillator, which may become a useful tool for precision measurements with squeezed light. The detector can be conceptually constructed out of two conventional heterodyne detectors (Fig. \ref{fig:pih2psh}). It turns out that the quantum nature of this detector cannot be satisfactorily described by the current theory of detection: The 3 dB noise penalty should take place due to the image sideband vacuum \cite{marino2007}. On the other hand, A 3 dB noise penalty should not take place in a phase-sensitive device, to be in consistency with the quantum theory of linear amplifier \cite{caves1982}.

%Otherwise, if the image sideband modes did contribute 3 dB heterodyne noise, a potential problem with the quantum theory of optical detection arises.
%On the other hand, however, The detector should be noise free at the quantum level because of its phase-sensitive nature \cite{caves1982}. %The dilemma reflects our lack of full understanding of the origin of the quantum noise in optical detection, especially the physics relevant to image sideband vacuum mode.

%A phase-sensitive heterodyne detector can be conceptually constructed out of two traditional heterodyne detectors in a way as illustrated in Fig. \ref{fig:pih2psh}. 

In this work, we develop a theory to describe the quantum behavior of the phase-sensitive detector, based on the quantum theory of optical coherence \cite{glauber1963b,ou1987,mandel1995}. We theoretically show that the studied detector is actually noise free, in a good agreement with experimental observation \cite{fan2014}, which may be explained by that the heterodyne detector measures only a single field of light at its input, according to our theoretical analysis. The results presented here are very important for a full understanding of the origin of the quantum noise in optical heterodyne detection.

In the next section, we summarize a relevant work \cite{ou1987,mandel1995} by Ou, Hong and Mandel, who studied certain two-time correlation functions of squeezed light and their Fourier conjugates. These functions are related to the auto-correlation functions and the spectral density of the photocurrent fluctuations that appear in homodyne measurements of the squeezed light. Needless to say, coherent light can be considered as a special kind of squeezed light when the degree of squeezing approaches zero. In Sec. III, we generalize the work by Ou, Hong and Mandel to the case of phase-sensitive heterodyne detection in the studied configuration. For coherent light as input to the detector, we calculate the noise figure (NF) of the detector. We discuss in Sec. IV the importance of the theoretical results obtained in this work. According to the theoretical analysis, the studied heterodyne detector owns both the advantages of a conventional heterodyne detector, which produces AC photoelectric signals avoiding DC noises, and its homodyne counterpart, which is noise free, showing the potential of being a useful tool for precision measurement of weak optical signals with non-classical light. As for the origin of the quantum noise in the detection, the heterodyne detector may measure a single field of light, i.e, the signal field. In this picture, the image sideband vacuum modes are part of the measured signal field, instead of belonging to other independent fields.%, which we propose an experiment with squeezed light to verify.

%significant difference of our theoretical development from the previous theory based on the concept of image sideband mode. 

\section{Fluctuations of the photoelectric current in homodyne measurement of light}

When treating the problem of homodyne measurement of light, Ou, Hong and Mandel considered a measured electromagnetic field $\hat{E}_s^{(+)}(\mathbf{r},t)$, whose effective bandwidth is nonzero but small compared with the midfrequency $\omega_s$, \cite{glauber1963b,ou1987}
\begin{equation}\label{eq:field}
\hat{E}_s^{(+)}(\mathbf{r},t)=\frac{i}{\sqrt{V}}\sum_{\mathbf{k}}\left(\frac{1}{2}\hbar\omega_{\mathbf{k}}\right)^{\frac{1}{2}}\hat{a}_{\mathbf{k}}e^{i(\mathbf{k}\cdot\mathbf{r}-\omega_{\mathbf{k}}t)},
\end{equation}
in which $V$ is the quantization volume and $\mathbf{k}$ stands for the set of plane-wave modes to which the detector responds with $\omega_{\mathbf{k}}$ the corresponding optical frequency of each mode. The amplitude operator $\hat{a}_{\mathbf{k}}$ is the photon annihilation operator for mode $\mathbf{k}$ and remains constant when there is no free electrical charge in the space \cite{glauber1963b}. The two mutually adjoint operators $\hat{a}_{\mathbf{k}}$ and $\hat{a}^\dagger_{\mathbf{k}}$ obey the following commutation relations
\begin{eqnarray}
[\hat{a}_{\mathbf{k}}, \hat{a}_{\mathbf{k'}}]=[\hat{a}^\dagger_{\mathbf{k}}, \hat{a}^\dagger_{\mathbf{k'}}] &=& 0, \nonumber \\  
\ [\hat{a}_{\mathbf{k}}, \hat{a}^\dagger_{\mathbf{k'}}]&=&\delta_{\mathbf{k},\mathbf{k'}}.
\end{eqnarray}

\begin{figure} 
\includegraphics[scale=0.30]{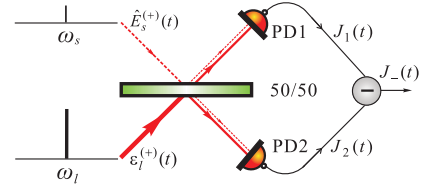}
\caption{\label{fig:hom} (color online) Homodyne measurement of light. The signal light $\hat{E}_s^{(+)}(t)$ interferes with the optical local oscillator $\mathscr{E}_l^{(+)}(t)$ at a balanced (50/50) beamsplitter. The mixed light at each output port of the beamsplitter is collected by a photodetector (PD1 or PD2) and the differenced photocurrent $J_-(t)\equiv J_1(t)-J_2(t)$ is sent to a spectrum analyzer for record. $\omega_s$ is the optical frequency of the excited center-frequency mode of the signal field, and $\omega_l$ is the frequency of the only excited mode of oscillator. For homodyne detection, $\omega_s=\omega_l$.}
\end{figure}

If the light field $\hat{E}_s^{(+)}(\mathbf{r},t)$ is detected in homodyne measurement as depicted in Fig. \ref{fig:hom}, the light intensities $\hat{I}_{1,2}(t)$ at the two output ports of the 50-50 beamsplitter read
\begin{eqnarray}\label{eq:intensity}
\hat{I}_{1,2}(t)&=& (1/2)\{ \hat{E}_s^{(-)}(t)\hat{E}_s^{(+)}(t) + \mathscr{E}_l^{(-)}(t)\mathscr{E}_l^{(+)}(t)\nonumber \\ 
&\pm& i[\mathscr{E}_l^{(+)}(t)\hat{E}_s^{(-)}(t)-\mathscr{E}_l^{(-)}(t)\hat{E}_s^{(+)}(t)]\}.
\end{eqnarray}
where $\mathscr{E}_l^{(+)}(t)=\mathscr{E}_le^{-i\omega_l t+i\theta_l}$ represents the classical single-frequency field of the local oscillator, with both the amplitude $\mathscr{E}_l$ and phase $\theta_l$ being real numbers. $\hat{E}_s^{(-)}(t)={\left[\hat{E}_s^{(+)}(t)\right]}^\dagger$ and $\mathscr{E}_l^{(-)}(t)={\left[\mathscr{E}_l^{(+)}(t)\right]}^*$. %For homodyne measurement, $\omega_s=\omega_l$. 

When the light falls on the photodiodes, it generates photoelectric emissions at certain times $t_1, t_2, ...$. If $j(t)$ is the current output pulse produced by a photoemission at time $t_i=0$, then the total photoelectric current $J(t)$ can be represented by the sum of pulses over all times $t_i$,
\begin{equation}
J(t)=\sum_i j(t-t_i).
\end{equation}
Apparently, $j(t-t_i)=0$ if $t<t_i$. Then, we obtain the power spectral density of the photocurrent fluctuations as \cite{ou1987,mandel1995}
\begin{equation}\label{eq:chi}
\chi(\omega)={\int}^{+\infty}_{-\infty}d\tau e^{i\omega \tau}<\Delta J_-(t) \Delta J_-(t+\tau)>_s,
\end{equation}
where $<\Delta J_-(t) \Delta J_-(t+\tau)>_s$ is the auto-correlation function of the differentiated-photocurrent fluctuations $\Delta J_-(t)$ ($\Delta A\equiv A-<A>_s$), with $<\cdot>_s$ standing for statistical averaging, produced by the two photodiodes. The function $<\Delta J_-(t) \Delta J_-(t+\tau)>_s$ can be grouped into two lumps, the auto-correlation functions of $\Delta J_i(t)$ ($i=1,2$) and the cross-correlation functions of $\Delta J_i(t)$ and $\Delta J_j(t)$ ($i,j=1,2$ and $i \ne j$), as follows:
\begin{eqnarray}\label{eq:autoJJ}
&&<\Delta J_-(t) \ \Delta J_-(t+\tau)>_s\nonumber \\
&=&\sum_{i=1,2}<\Delta J_i(t) \Delta J_i(t+\tau)>_s\nonumber \\
&-&\sum_{i,j=1,2;i\ne j}<\Delta J_i(t) \Delta J_j(t+\tau)>_s,
\end{eqnarray}
wherein the auto-correlation functions of $\Delta J_i(t)$ ($i=1,2$) read \cite{ou1987,mandel1995,feng2012}
\begin{eqnarray}\label{eq:autoJ}
& &<\Delta J_i(t) \Delta J_i(t+\tau)>_s  \ \ (i=1,2)\nonumber\\
&=&\eta \int^{\infty}_0 dt'<\hat{I}_i(t-t')>j_i(t')j_i(t'+\tau)\nonumber\\
&+&\eta^2 \int\!\!\!\int_0^{\infty}dt'dt''\lambda_i(t-t',\tau+t'-t'')j_i(t')j_i(t'').
\end{eqnarray}
Here we assumed two identical photodiodes that are characterized by the same parameter $\eta$ for their quantum efficiency, and $j_i(t-t')$ ($i=1,2$) are the photoelectrical current pulses produced in the $i$th photodiodes for $t>t'$ \cite{ou1987,mandel1995}. $\lambda_i(t,\iota)\equiv<\mathscr{T}:\Delta \hat{I}_i(t)\Delta \hat{I}_i(t+\iota):>$ ($i=1,2$) are the auto-correlation functions of light-intensity fluctuations and the symbol $\mathscr{T}: :$ means time- and normal-ordering of the field operators $\hat{E}_s^{(\pm)}(t)$. %One should note that included in the signal-field operator (\ref{eq:field}) are all possible modes of frequency \cite{mandel1995,glauber1963b}, whether they are excited or not. 
The first term in Eq. (\ref{eq:autoJ}) accounts for the shot noise of light, while the second one depends on the fluctuation nature of the light being detected. 

The auto-correlation functions $\lambda_i(t,\iota)$ ($i=1,2$) can be calculated as (for detailed calculations, refer to the appendix of \cite{feng2012} with the only difference being that the oscillator is bichromatic and the beamsplitter is extremely unbalanced there)
\begin{eqnarray}\label{eq:app4}
\lambda_i(t,\iota)&=&(1/4)\times\nonumber\\ 
&&\big[\mathscr{E}^{(+)}_l(t)\mathscr{E}^{(-)}_l(t+\iota)<\Delta\hat{E}_s^{(-)}(t)\Delta\hat{E}_s^{(+)}(t+\iota)> \nonumber \\
&+&\mathscr{E}^{(-)}_l(t)\mathscr{E}^{(+)}_l(t+\iota)<\Delta\hat{E}_s^{(-)}(t+\iota)\Delta\hat{E}_s^{(+)}(t)> \nonumber \\
&-&\mathscr{E}^{(+)}_l(t)\mathscr{E}^{(+)}_l(t+\iota)<\Delta\hat{E}_s^{(-)}(t)\Delta\hat{E}_s^{(-)}(t+\iota)> \nonumber \\
&-&\mathscr{E}^{(-)}_l(t)\mathscr{E}^{(-)}_l(t+\iota)<\Delta\hat{E}_s^{(+)}(t+\iota)\Delta\hat{E}_s^{(+)}(t)>\big] \nonumber \\
&+& O(\mathscr{E}_l). \ \ 
\end{eqnarray}
For the sake of simplicity, we only consider coherent light here. According to the definition of coherent states \cite{glauber1963b}, which are eigenstates of the field operator $\hat{E}_s^{(+)}(t)$, all the second-order correlation functions of the field fluctuations in the above equation vanish, i.e., $\lambda_i(t,\iota)\approx 0$. %One must note that the aforesaid image-band mode was already included in $\hat{E}_s^{(-)}(t)$ in the above calculations for optical heterodyning! 
With this in mind and plugging Eq. (\ref{eq:intensity}) into Eq. (\ref{eq:autoJ}), we arrive at
\begin{eqnarray}\label{eq:autoJ1}
& &<\Delta J_i(t) \Delta J_i(t+\tau)>_s= \frac{\eta\mathscr{E}_l^2}{2} \int^{\infty}_0 dt'j_i(t')j_i(t'+\tau).\ \ 
\end{eqnarray}
In the mathematical manipulations, only the leading term $\mathscr{E}_l^{(-)}(t)\mathscr{E}_l^{(+)}(t)=\mathscr{E}_l^2$ in Eq. (\ref{eq:intensity}) remained under the strong-oscillator approximation.

Similarly, one can show without much difficulty that the cross-correlation functions $<\Delta J_i(t) \Delta J_j(t+\tau)>_s= 0$ for $i\ne j$ ($i,j=1,2$). Therefore, with Eq. (\ref{eq:autoJJ}), we obtain
\begin{eqnarray}\label{eq:autoJm}
& &<\Delta J_-(t) \Delta J_-(t+\tau)>_s= \eta\mathscr{E}_l^2 \int^{\infty}_0 dt'j(t')j(t'+\tau),\ \ \ 
\end{eqnarray}
in which two identical photodiodes were assumed again for the detector, i.e., $j(t)\equiv j_1(t)=j_2(t)$.

Now let $K(\omega)$ be the Fourier transform of the photocurrent pulses $j(t)$,
\begin{equation}\label{eq:komega}
K(\omega)= \int_0^{\infty}d\tau j(\tau)e^{i\omega\tau},
\end{equation}
which can be interpreted as the frequency response of both photodiodes. Substituting Eqs. (\ref{eq:autoJm}) and (\ref{eq:komega}) into Eq. (\ref{eq:chi}) leads to \cite{ou1987,mandel1995}
\begin{equation}\label{eq:chi1}
\chi(\omega)=2\eta\mathscr{E}_l^2 |K(\omega)|^2.
\end{equation}
The factor of two here accounts for the contribution of negative-frequency components when the calculation is compared with practical measurement. For ideal photodiodes with sufficient response speeds, the photocurrent pulses may be approximated by Delta functions. Under this approximation, $|K(\omega)|=e$, the charge on the electron. Then, we have the power spectral density of the photocurrent fluctuations
\begin{equation}\label{eq:chi2}
\chi(\omega)=2\eta e^2 \mathscr{E}_l^2.
\end{equation}

\section{Quantum noise in heterodyne measurement of light with a bichromatic local oscillator}

\begin{figure} 
\includegraphics[scale=0.28]{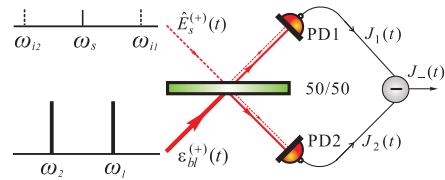}
\caption{\label{fig:het} (color online) Heterodyne measurement of light with a bichromatic local oscillator. The signal light $\hat{E}_s^{(+)}(t)$ interferes with the bichromatic local oscillator $\mathscr{E}_{bl}^{(+)}(t)$ at a balanced (50/50) beamsplitter. The mixed light at each output port of the beamsplitter is collected by a photodetector (PD1 or PD2) and the differenced photocurrent $J_-(t)\equiv J_1(t)-J_2(t)$ is sent to a spectrum analyzer for record. The local oscillator consists of two excited modes of frequency, respectively, at $\omega_1$ and $\omega_2$. Two image sideband vacuum modes at $\omega_{i1}=\omega_1+\Omega$ and $\omega_{i2}=\omega_2-\Omega$ are shown together with the signal mode. $\Omega\equiv\omega_1-\omega_s=\omega_s-\omega_2$ is the heterodyne frequency.}
\end{figure}

Now we consider the same electromagnetic field  $\hat{E}_s^{(+)}(\mathbf{r},t)$ to be measured by a heterodyne detector with a bichromatic local oscillator (Fig. \ref{fig:het}),
\begin{equation}\label{eq:bilocal}
\mathscr{E}_{bl}^{(+)}(t)=(\mathscr{E}_l/\sqrt{2})\left(e^{-i\omega_{1} t+i\theta_{1}}+e^{-i\omega_{2} t+i\theta_{2}}\right).
\end{equation}
One should note that the treatment of Ou, Hong and Mandel is based on the stationary-photocurrent assumption \cite{ou1987,mandel1995}, which is not valid any more in heterodyne measurement. We take advantage of the fact that the practical RBW, $\Omega_r$, of the spectral analyzer must be set to satisfy $\Omega_r<<\Omega$, the heterodyne frequency, to resolve the beatnote signal in the photocurrent. In other words, the heterodyne measurement must be accomplished during a period of time $T\sim{\Omega_r}^{-1}>>{\Omega}^{-1}$ or longer. Consequently, to treat the heterodyne problem, we only need to assume that the average photocurrent over the period of time $T\sim{\Omega_r}^{-1}$ is stationary. Under this assumption, we concern the average power spectral density of the photocurrent fluctuations (see Ref. \cite{collett1987,drummond90} for similar treatments),
\begin{equation}\label{eq:chihet}
\chi(\omega)=\frac{1}{T}{\int}^{T}_{0}dt{\int}^{+\infty}_{-\infty}d\tau e^{i\omega \tau}<\Delta J_-(t) \Delta J_-(t+\tau)>_s.
\end{equation}

\begin{figure} 
\includegraphics[scale=0.35]{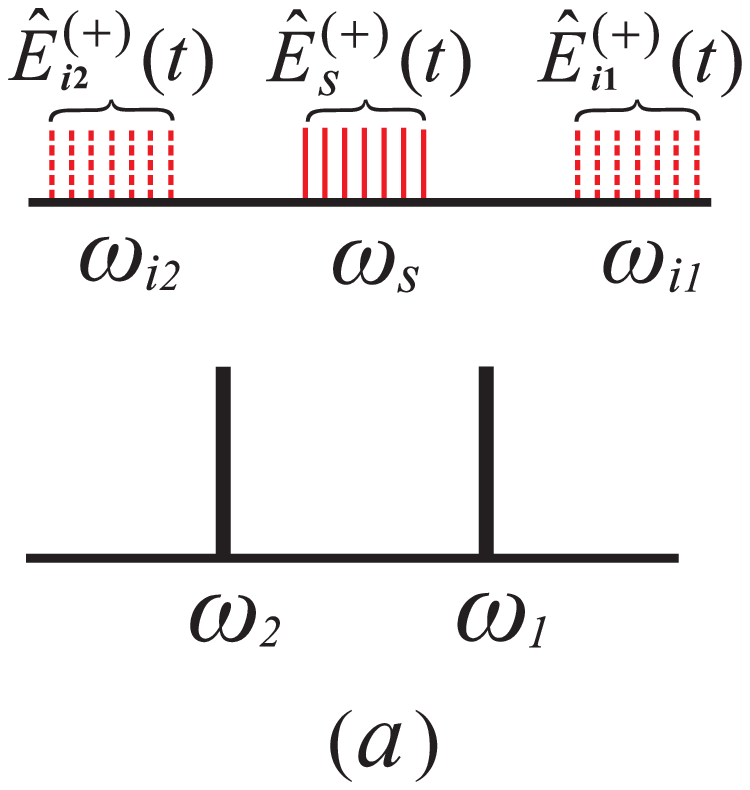}\hspace{0.6in}
\includegraphics[scale=0.35]{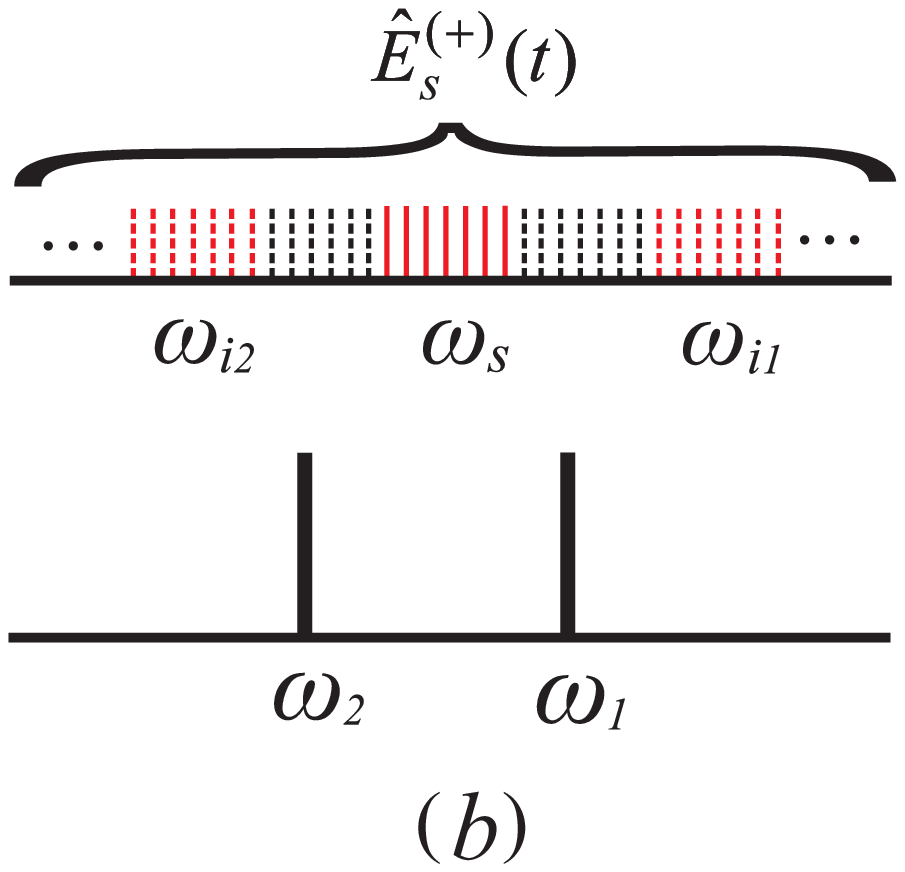}
\caption{\label{fig:field} (color online) Illustration of two different ways to treat the input light: (a) The detector measures three independent fields and the image sideband vacuum modes belong to two image sideband fields, $\hat{E}_{i1}^{(+)}(t)$ and $\hat{E}_{i2}^{(+)}(t)$, respectively. (b) The detector measures only one light field, i.e., the signal field $\hat{E}_s^{(+)}(\mathbf{r},t)$, and the image sideband vacuum modes are part of the signal light field. }
\end{figure}

Now there are two different ways to treat the signal field $\hat{E}_s^{(+)}(\mathbf{r},t)$: (1) The image sideband vacuum modes are not within the bandwidth of the signal field and involved in the detection are there two more independent fields $\hat{E}_{i1}^{(+)}(\mathbf{r},t)$ and $\hat{E}_{i2}^{(+)}(\mathbf{r},t)$, namely image sideband fields \cite{yuen1980},
\begin{equation}\label{eq:imagefields}
\hat{E}_{i1,i2}^{(+)}(\mathbf{r},t)=\frac{i}{\sqrt{V}}\sum_{\mathbf{k}}\left(\frac{1}{2}\hbar\omega_{\mathbf{k}}\right)^{\frac{1}{2}}\hat{b}_{\mathbf{k}}e^{i(\mathbf{k}\cdot\mathbf{r}-\omega_{\mathbf{k}}t)},
\end{equation}
whose center-frequency modes are located at $\omega_{i1}$ and $\omega_{i2}$, respectively (Fig. \ref{fig:field}a). Here the amplitude operator $\hat{b}_{\mathbf{k}}$ is the photon annihilation operator for mode $\mathbf{k}$. Or (2) the image sideband vacuum modes are part of the field $\hat{E}_s^{(+)}(\mathbf{r},t)$, which is the only field that the heterodyne detector measures (Fig. \ref{fig:field}b).

In any case, the auto-correlation function of the differenced-photocurrent fluctuations reads
\begin{eqnarray}\label{eq:autoJJhet}
&&<\Delta J_-(t) \ \Delta J_-(t+\tau)>_s\nonumber \\
&=&\sum_{i=1}^2\eta \int^{\infty}_0 dt'<\hat{I}_i(t-t')>j_i(t')j_i(t'+\tau)\nonumber\\
&+&\sum_{i,j=1}^2 \eta^2(-1)^{i+j} \int\!\!\!\int_0^{\infty}dt'dt''j_i(t')j_j(t'')\nonumber \\
&\times& \lambda_{ij}(t-t',\tau+t'-t''),
\end{eqnarray}
in which $\lambda_{ij}(t,\iota)\equiv<\mathscr{T}:\Delta \hat{I}_i(t)\Delta \hat{I}_j(t+\iota):>$ ($i,j=1,2$) are the correlation functions of light-intensity fluctuations. In the heterodyne case, the light intensities $\hat{I}_{1,2}(t)$ at the two output ports of the 50-50 beamsplitter read
\begin{eqnarray}\label{eq:intensityhet}
\hat{I}_{1,2}(t)&=& (1/2)\{ \hat{E}_t^{(-)}(t)\hat{E}_t^{(+)}(t) + \mathscr{E}_{bl}^{(-)}(t)\mathscr{E}_{bl}^{(+)}(t)\nonumber \\ 
&\pm& i[\mathscr{E}_{bl}^{(+)}(t)\hat{E}_t^{(-)}(t)-\mathscr{E}_{bl}^{(-)}(t)\hat{E}_t^{(+)}(t)]\},
\end{eqnarray}
wherein $\hat{E}_t^{(+)}(t)$ stands for the input light field and 
\begin{equation} \label{eq:multifield}
\hat{E}_t^{(+)}(t)=\hat{E}_s^{(+)}(t)+\hat{E}_{i1}^{(+)}(t)+\hat{E}_{i2}^{(+)}(t),
\end{equation}
if the detector senses two more independent fields in addition to the signal field $\hat{E}_s^{(+)}(t)$ (Fig. \ref{fig:field}a). Otherwise, if the detector measures only one field, i.e., the signal field (Fig. \ref{fig:field}b),
\begin{equation} \label{eq:onefield}
\hat{E}_t^{(+)}(t)=\hat{E}_s^{(+)}(t).
\end{equation}

Then we obtain the correlation functions of the photocurrent fluctuations as follows
\begin{eqnarray}\label{eq:lambdahet}
&&\lambda_{ij}(t,\iota)=\nonumber\\ 
&&\big[\mathscr{E}^{(+)}_{bl}(t)\mathscr{E}^{(-)}_{bl}(t+\iota)<\Delta\hat{E}_t^{(-)}(t)\Delta\hat{E}_t^{(+)}(t+\iota)> \nonumber \\
&+&\mathscr{E}^{(-)}_{bl}(t)\mathscr{E}^{(+)}_{bl}(t+\iota)<\Delta\hat{E}_t^{(-)}(t+\iota)\Delta\hat{E}_t^{(+)}(t)> \nonumber \\
&-&\mathscr{E}^{(+)}_{bl}(t)\mathscr{E}^{(+)}_{bl}(t+\iota)<\Delta\hat{E}_t^{(-)}(t)\Delta\hat{E}_t^{(-)}(t+\iota)> \nonumber \\
&-&\mathscr{E}^{(-)}_{bl}(t)\mathscr{E}^{(-)}_{bl}(t+\iota)<\Delta\hat{E}_t^{(+)}(t+\iota)\Delta\hat{E}_t^{(+)}(t)>\big]/4 \nonumber \\
&+& O(\mathscr{E}_l). \ \ 
\end{eqnarray}

For coherent light at the input of the detector, the field $\hat{E}_s^{(+)}(t)$ is in a coherent state and the image sideband fields $\hat{E}_{i1}^{(+)}(t)$ and $\hat{E}_{i2}^{(+)}(t)$, if any, are in vacuum states. Then, the correlation functions $\lambda_{ij}(t,\iota)\approx 0$ no matter what we choose between Eq. (\ref{eq:multifield}) and Eq. (\ref{eq:onefield}), according to the definition of coherent state \cite{glauber1963b}. Consequently, with Eqs. (\ref{eq:chihet}), (\ref{eq:autoJJhet}) and (\ref{eq:intensityhet}), we have
\begin{eqnarray}\label{eq:chihet1}
\chi(\omega)&=&\frac{1}{T}{\int}^{T}_{0}dt{\int}^{+\infty}_{-\infty}d\tau e^{i\omega \tau}\nonumber \\
&\times&\sum_{i=1}^2 \eta\int^{\infty}_0 dt'<\hat{I}_i(t-t')>j_i(t')j_i(t'+\tau)\nonumber \\
&=& \eta \int^{\infty}_0 dt'{\int}^{+\infty}_{-\infty}d\tau e^{i\omega \tau}j(t')j(t'+\tau) \nonumber \\
&\times&\frac{1}{T}{\int}^{T}_{0}dt\mathscr{E}_{bl}^{(-)}(t)\mathscr{E}_{bl}^{(+)}(t),
\end{eqnarray}
in which, as in the homodyne case, we assumed two identical photodiodes in the last step for the detector and only the leading terms remained in the calculation. Plugging Eqs. (\ref{eq:komega}) and (\ref{eq:bilocal}) into Eq. (\ref{eq:chihet1}) leads to
\begin{eqnarray}\label{eq:chihet2}
\chi(\omega)&=&2\eta \mathscr{E}_l^2 |K(\omega)|^2.
\end{eqnarray}
Again the factor of two comes from the contribution of negative-frequency components. Under the same approximation that the photodiodes have sufficient response speeds, $|K(\omega)|=e$, we have the power spectral density for the photocurrent fluctuations
\begin{equation}\label{eq:chihet3}
\chi(\omega)=2\eta e^2 \mathscr{E}_l^2,
\end{equation}
which is identical to Eq. (\ref{eq:chi2}).

Next, we consider the average power of the output photoelectrical signal produced by the heterodyne detector. The average photocurrent signal at the output of the detector reads \cite{ou1987,mandel1995,feng2012}
\begin{eqnarray}
&&<J_-(t)>\nonumber\\
&=&\sum_i <j_1(t-t_i)>-\sum_i <j_2(t-t_i)>\nonumber\\
&=&\sum_i j(t-t_i) P_1(t_i)\Delta t_i-\sum_i j(t-t_i) P_2(t_i)\Delta t_i,\nonumber
\end{eqnarray}
in which $P_i(t)\Delta t = \eta<\hat{I}_i(t)> \Delta t$ ($i=1,2$) is the probability of photodetection registered by the $i$th photodiode at time $t$ within time interval $\Delta t$ \cite{glauber1963a,glauber1963b}. Converting the summation into integrand, we arrive at
\begin{eqnarray}
<J_-(t)>=\eta \int^{\infty}_0 dt' j(t')<\hat{I}_1(t-t')-\hat{I}_2(t-t')>.\nonumber
\end{eqnarray}
For simplicity without loss of generality, we assume unity load resistance for the spectrum analyzer and that the signal light has only a single excited mode of frequency in a coherent state. One should note that we previously assumed photodiodes with sufficient response speeds, $j(t)=e\delta(t)$. Then the output photoelectrical signal reads
\begin{eqnarray}
&&<J_-(t)>\nonumber\\
&=&\eta e <\hat{I}_1(t)-\hat{I}_2(t)>\nonumber\\
&=&i \eta e \left[\mathscr{E}_{bl}^{(+)}(t)<\hat{E}_t^{(-)}(t)>-\mathscr{E}_{bl}^{(-)}(t)<\hat{E}_t^{(+)}(t)>\right].\nonumber
\end{eqnarray}
It is not difficult to show 
\begin{eqnarray}
<\hat{E}_t^{(+)}(t)>&=&\frac{i}{\sqrt{2V}}\alpha_{\mathbf{k}_s}e^{i(\mathbf{k}_s\cdot\mathbf{r}-\omega_{\mathbf{k}_s}t)}\nonumber \\
<\hat{E}_t^{(-)}(t)>&=&-\frac{i}{\sqrt{2V}}\alpha_{\mathbf{k}_s}e^{-i(\mathbf{k}_s\cdot\mathbf{r}-\omega_{\mathbf{k}_s}t)}.\nonumber
\end{eqnarray}
Here $\alpha_{\mathbf{k}_s}=<\hat{a}_{\mathbf{k}_s}>\sqrt{\hbar\omega_{\mathbf{k}_s}}$, and $\mathbf{k}_s$ stands for the excited mode of frequency in the signal field. With Eq. (\ref{eq:bilocal}), one can rewrite the output photoelectrical signal as
\begin{eqnarray}\label{eq:Jave}
&&<J_-(t)>\nonumber\\
&=&\frac{\eta e \alpha_s \mathscr{E}_{l}}{2\sqrt{V}}\left[e^{-i\omega_st+i\theta_s}\left(e^{i\omega_{1}t-i\theta_{1}}+e^{i\omega_{2}t-i\theta_{2}}\right)+c.c.\right]\nonumber\\
&=&\frac{2\eta e \alpha_s \mathscr{E}_{l}}{\sqrt{V}}\cos(\theta_s-\bar{\theta})\cos(\Omega t + \Delta \theta),
\end{eqnarray}
in which $\omega_s\equiv\omega_{\mathbf{k}_s}$, $\alpha_s\equiv\alpha_{\mathbf{k}_s}$, $\theta_s=\mathbf{k}_s\cdot\mathbf{r}$, $\bar{\theta}=(\theta_1+\theta_2)/2$, and $\Delta\theta=\theta_2-\theta_1$. Accordingly, the average power of the output signal is
\begin{eqnarray}\label{eq:pave}
&&P_{out}=\frac{1}{T}\int_0^Tdt <<J_-(t)>^2>_s\nonumber\\
&=&\frac{4(\eta e \alpha_s \mathscr{E}_{l})^2}{TV}\int_0^Tdt \cos^2(\Omega t + \Delta \theta)<\cos^2(\theta_s-\bar{\theta})>_s\nonumber\\
&=&\frac{(\eta e \alpha_s \mathscr{E}_{l})^2}{V}.
\end{eqnarray}
%Since the output signal has two components of frequency, of which only the component of positive frequency can be detected by the spectrum analyzer. Therefore, the measured average power of the output signal is half of that in Eq. (\ref{eq:pave}),
%\begin{eqnarray}\label{eq:pave1}
%P_{out}=\frac{(\eta e \alpha_s \mathscr{E}_{l})^2}{2V}.
%\end{eqnarray}
With Eqs. (\ref{eq:chihet3}) and (\ref{eq:pave}), we calculate the SNR at the output of the heterodyne detector as, for one second of measurement time,
\begin{eqnarray}\label{eq:snrout}
\mbox{SNR}_{out}&=&\frac{P_{out}}{\chi(\omega)}=\frac{(\eta e \alpha_s \mathscr{E}_{l})^2}{2V\eta e^2 \mathscr{E}_l^2}=\frac{\eta \alpha_s^2 }{2V}.
\end{eqnarray}
Concerning the SNR at the input of the heterodyne detector, it is the inherent SNR of input signal light, which is in a coherent state. For one second of measurement time, we have
\begin{eqnarray}\label{eq:snrin}
\mbox{SNR}_{in}&=&\frac{<\hat{I}(t)>^2}{<\hat{I}(t)>}=<\hat{I}(t)>\nonumber \\
&=&<\hat{E}_t^{(-)}(t)\hat{E}_t^{(+)}(t)>=\frac{\eta \alpha_s^2 }{2V}.
\end{eqnarray}
Finally, we obtain the noise figure of the heterodyne phase-sensitive detector
\begin{eqnarray}\label{eq:nf}
\mbox{NF}=10\log_{10}\frac{\mbox{SNR}_{in}}{\mbox{SNR}_{out}}=0 \ \mbox{dB}
\end{eqnarray}

\section{Discussions}

We have shown that, with coherent light at the input, a phase-sensitive heterodyne detector with a bichromatic local oscillator is noise free at the quantum level. This is in good agreement with the quantum theory of linear amplifier \cite{caves1982}. Therefore, as a prospective tool for precision measurements with squeezed light, the studied heterodyne detector has both advantages of a conventional heterodyne detector, where classical noises around DC area are avoided, and its homodyne counterpart, which is noise free at the quantum level.

As for the origin of the quantum noise, one must notice that the result presented here remains no matter if one takes into account the image sideband fields in the calculation, i.e., the result is independent of the choice between Eq. (\ref{eq:multifield}) and Eq. (\ref{eq:onefield}). In this sense, we cannot tell if the quantum noise in the heterodyne detector originates from three independent fields, as described by Eq. (\ref{eq:multifield}), or from a single field of light, as delineated by Eq. (\ref{eq:onefield}).

Nonetheless, the power spectral density for the photocurrent fluctuations produced by the heterodyne detector, Eq. (\ref{eq:chihet3}), is identical to that by a homodyne detector, Eq. (\ref{eq:chi2}). It is well known that the homodyne quantum noise results from only one field in the detection. From this, it is reasonable to speculate that the quantum noise represented by Eq. (\ref{eq:chihet3}) originates from only one field too. If this is true, all the image sideband vacuum modes must be treated as part of the signal field $\hat{E}_s^{(+)}(\mathbf{r},t)$ with Eq. (\ref{eq:onefield}).

In a recently accomplished work, we discovered \cite{fan2014} that the 3 dB extra quantum noise due to the image sideband vacuum modes was indeed absent, or significantly suppressed, in experimental observation, in good agreement with our theoretical analysis (Table \ref{tab:nf}). However, the mechanism for the absence of the 3 dB heterodyne noise was unknown.

\begin{table}%[H] add [H] placement to break table across pages
\caption{\label{tab:nf} Noise figure of the studied phase-sensitive heterodyne detector, as calculated according to Eq. (\ref{eq:nf}), for coherent light at the input. To enable a direct comparison with the experimental results \cite{fan2014}, we provide here the calculation results for signal light at the power levels of 0.5 nW, 1.0 nW, and 2.0 nW, respectively. The SNR at the input of the detector is computed as SNR$_{in}=10\log_{10}\bar{N}$, where $\bar{N}$ is the number of detected photons within a period of 1 ms. A quantum efficiency of $\eta=70\%$ is assumed for the detector. SNR$_{out}$ is the SNR of the photoelectric signal, according to Eq. (\ref{eq:snrout}), integrated over all the detection volume $V$, with RBW = 1kHz.
}
\begin{ruledtabular}
\begin{tabular}{c|c|c|c}
%Lines of table here ending with \\
\ \ \ \ $P_{s} $ (nW) \ \ \ & SNR$_{in}$ (dB)\ \ \ \ & SNR$_{out}$ (dB) \ \ \ \ & NF (dB) \\ \hline
0.5 & 62.68  \ \ \ \ & 62.68 \ \ \ \ & 0.00 \\\hline
1.0 & 65.69  \ \ \ \ & 65.69 \ \ \ \ & 0.00 \\\hline
2.0 & 68.70  \ \ \ \ & 68.70 \ \ \ \ & 0.00 \\
\end{tabular}
\end{ruledtabular}
\end{table}

One might explain the experimental observation with a destructive quantum interference between the two image sideband modes that caused a cancel of their quantum noises in the detection. Nevertheless, this entails quantum anti-correlation between the two vacuum modes, which sounds very unlikely for coherent light. Now the experimental results may be explained in the following picture: The phase-sensitive heterodyne detector actually measured only one field of light at its input, and it was this measured signal field that produced the quantum noise in the heterodyne detection, the same as in homodyne detection.

To verify the above speculation, one may utilize an optical signal in a squeezed state at the input of the heterodyne detector. For squeezed light under measurement, the correlation functions $\lambda_{ij}(t,\iota)$ are nonzero in Eq. (\ref{eq:autoJJhet}).
%\begin{eqnarray}\label{eq:autoJJhet2}
%&&<\Delta J_-(t) \ \Delta J_-(t+\tau)>_s\nonumber \\
%&=&\sum_{i=1}^2\eta \int^{\infty}_0 dt'<\hat{I}_i(t-t')>j_i(t')j_i(t'+\tau) \nonumber \\
%&+&\sum_{i,j=1}^2 \eta^2(-1)^{i+j} \nonumber \\
%&\times&\int\!\!\!\int_0^{\infty}dt'dt''j_i(t')j_j(t'')\lambda_{ij}(t-t',\tau+t'-t'').
%\end{eqnarray}
From Eq. (\ref{eq:lambdahet}), one can easily see that the correlation functions $\lambda_{ij}(t,\iota)$ may be different for the two cases of Eqs. (\ref{eq:multifield}) and (\ref{eq:onefield}). 

If the two image sideband fields are within the squeezing bandwidth of the signal field, they should be quantum correlated and make observable contributions to the power spectral density $\chi(\omega)$. By experimentally measuring $\chi(\omega)$ and comparing the result with theoretical expectations respectively based on Eqs. (\ref{eq:multifield}) and (\ref{eq:onefield}), one may know the source of the quantum noise in the phase-sensitive heterodyne detector. 

%to account for the experimental observation, one has to assume that the two image sideband vacuum modes involved in the measurement contributed no extra quantum noise \cite{fan2014}. This assumption seems too artificial to some extent and needs justification. Moreover, further investigation is highly demanded to reveal the physics of the disappearance, or significant suppression, of the expected 3 dB extra quantum noise in the heterodyne detection.

%The theoretical development, in good agreement with experiment, may provide us a picture about the origin of the quantum noise in optical detection that is different from the picture based on the concept of image sideband mode. An experiment with squeezed light is proposed to verify % As one will see below, the theory developed here, in good agreement with experiment, can naturally circumvent the theoretical problem discussed above, whereas the existing theory based on the concept of image sideband mode does not \cite{fan2014}.

\section{Conclusions}

We have developed a theory to describe the quantum behavior of a phase-sensitive heterodyne detector with a bichromatic local oscillator, assuming coherent light at the input. We have shown, from the viewpoint of theoretical analysis, that the studied heterodyne detector is noise free at the quantum level, in a good agreement with experiment, indicating that the detector has the potential of becoming a useful tool for precision measurement with squeezed light. We have put forth a new picture, for the origin of the quantum noise in the heterodyne detector, that the detector measures a single field of light at its input, with all the image sideband vacuum modes being part of the signal field.% We propose that, by utilizing squeezed light as input to the detector, an experiment may be performed to verify whether the quantum noise originates from a single field of light, or from multi-field of light.

\begin{acknowledgments}
This work was supported by the National Natural Science Foundation of China (grant No. 11174094).
\end{acknowledgments}

\bibliography{noiselesspia}

\providecommand{\noopsort}[1]{}\providecommand{\singleletter}[1]{#1}%
\begin{thebibliography}{21}%
\makeatletter
\providecommand \@ifxundefined [1]{%
 \@ifx{#1\undefined}
}%
\providecommand \@ifnum [1]{%
 \ifnum #1\expandafter \@firstoftwo
 \else \expandafter \@secondoftwo
 \fi
}%
\providecommand \@ifx [1]{%
 \ifx #1\expandafter \@firstoftwo
 \else \expandafter \@secondoftwo
 \fi
}%
\providecommand \natexlab [1]{#1}%
\providecommand \enquote  [1]{``#1''}%
\providecommand \bibnamefont  [1]{#1}%
\providecommand \bibfnamefont [1]{#1}%
\providecommand \citenamefont [1]{#1}%
\providecommand \href@noop [0]{\@secondoftwo}%
\providecommand \href [0]{\begingroup \@sanitize@url \@href}%
\providecommand \@href[1]{\@@startlink{#1}\@@href}%
\providecommand \@@href[1]{\endgroup#1\@@endlink}%
\providecommand \@sanitize@url [0]{\catcode `\\12\catcode `\$12\catcode
  `\&12\catcode `\#12\catcode `\^12\catcode `\_12\catcode `\%12\relax}%
\providecommand \@@startlink[1]{}%
\providecommand \@@endlink[0]{}%
\providecommand \url  [0]{\begingroup\@sanitize@url \@url }%
\providecommand \@url [1]{\endgroup\@href {#1}{\urlprefix }}%
\providecommand \urlprefix  [0]{URL }%
\providecommand \Eprint [0]{\href }%
\providecommand \doibase [0]{http://dx.doi.org/}%
\providecommand \selectlanguage [0]{\@gobble}%
\providecommand \bibinfo  [0]{\@secondoftwo}%
\providecommand \bibfield  [0]{\@secondoftwo}%
\providecommand \translation [1]{[#1]}%
\providecommand \BibitemOpen [0]{}%
\providecommand \bibitemStop [0]{}%
\providecommand \bibitemNoStop [0]{.\EOS\space}%
\providecommand \EOS [0]{\spacefactor3000\relax}%
\providecommand \BibitemShut  [1]{\csname bibitem#1\endcsname}%
\let\auto@bib@innerbib\@empty
%</preamble>
\bibitem [{\citenamefont {Personick}(1971)}]{personick1971}%
  \BibitemOpen
  \bibfield  {author} {\bibinfo {author} {\bibfnamefont {S.~D.}\ \bibnamefont
  {Personick}},\ }\href@noop {} {\bibfield  {journal} {\bibinfo  {journal}
  {Bell Syst. Tech. J.}\ }\textbf {\bibinfo {volume} {50}},\ \bibinfo {pages}
  {213} (\bibinfo {year} {1971})}\BibitemShut {NoStop}%
\bibitem [{\citenamefont {Yuen}\ and\ \citenamefont
  {Shapiro}(1978)}]{yuen1978}%
  \BibitemOpen
  \bibfield  {author} {\bibinfo {author} {\bibfnamefont {H.~P.}\ \bibnamefont
  {Yuen}}\ and\ \bibinfo {author} {\bibfnamefont {J.~H.}\ \bibnamefont
  {Shapiro}},\ }\href@noop {} {\bibfield  {journal} {\bibinfo  {journal} {IEEE
  Transactions on Information Theory}\ }\textbf {\bibinfo {volume} {24}},\
  \bibinfo {pages} {657} (\bibinfo {year} {1978})}\BibitemShut {NoStop}%
\bibitem [{\citenamefont {Shapiro}\ \emph {et~al.}(1979)\citenamefont
  {Shapiro}, \citenamefont {Yuen},\ and\ \citenamefont {Mata}}]{shapiro1979}%
  \BibitemOpen
  \bibfield  {author} {\bibinfo {author} {\bibfnamefont {J.~H.}\ \bibnamefont
  {Shapiro}}, \bibinfo {author} {\bibfnamefont {H.~P.}\ \bibnamefont {Yuen}}, \
  and\ \bibinfo {author} {\bibfnamefont {J.~A.~M.}\ \bibnamefont {Mata}},\
  }\href@noop {} {\bibfield  {journal} {\bibinfo  {journal} {IEEE Transactions
  on Information Theory}\ }\textbf {\bibinfo {volume} {25}},\ \bibinfo {pages}
  {179} (\bibinfo {year} {1979})}\BibitemShut {NoStop}%
\bibitem [{\citenamefont {Yuen}\ and\ \citenamefont
  {Shapiro}(1980)}]{yuen1980}%
  \BibitemOpen
  \bibfield  {author} {\bibinfo {author} {\bibfnamefont {H.~P.}\ \bibnamefont
  {Yuen}}\ and\ \bibinfo {author} {\bibfnamefont {J.~H.}\ \bibnamefont
  {Shapiro}},\ }\href@noop {} {\bibfield  {journal} {\bibinfo  {journal} {IEEE
  Transactions on Information Theory}\ }\textbf {\bibinfo {volume} {26}},\
  \bibinfo {pages} {78} (\bibinfo {year} {1980})}\BibitemShut {NoStop}%
\bibitem [{\citenamefont {Yuen}\ and\ \citenamefont {Chan}(1983)}]{yuen1983}%
  \BibitemOpen
  \bibfield  {author} {\bibinfo {author} {\bibfnamefont {H.~P.}\ \bibnamefont
  {Yuen}}\ and\ \bibinfo {author} {\bibfnamefont {V.~W.~S.}\ \bibnamefont
  {Chan}},\ }\href@noop {} {\bibfield  {journal} {\bibinfo  {journal} {Opt.
  Lett.}\ }\textbf {\bibinfo {volume} {8}},\ \bibinfo {pages} {177} (\bibinfo
  {year} {1983})}\BibitemShut {NoStop}%
\bibitem [{\citenamefont {Schumaker}(1984)}]{schumaker1984}%
  \BibitemOpen
  \bibfield  {author} {\bibinfo {author} {\bibfnamefont {B.~L.}\ \bibnamefont
  {Schumaker}},\ }\href@noop {} {\bibfield  {journal} {\bibinfo  {journal}
  {Opt. Lett.}\ }\textbf {\bibinfo {volume} {5}},\ \bibinfo {pages} {189}
  (\bibinfo {year} {1984})}\BibitemShut {NoStop}%
\bibitem [{\citenamefont {Yamamoto}\ and\ \citenamefont
  {Haus}(1986)}]{yamamoto1986}%
  \BibitemOpen
  \bibfield  {author} {\bibinfo {author} {\bibfnamefont {Y.}~\bibnamefont
  {Yamamoto}}\ and\ \bibinfo {author} {\bibfnamefont {H.~A.}\ \bibnamefont
  {Haus}},\ }\href@noop {} {\bibfield  {journal} {\bibinfo  {journal} {Rev.
  Mod. Phys.}\ }\textbf {\bibinfo {volume} {58}},\ \bibinfo {pages} {1001}
  (\bibinfo {year} {1986})}\BibitemShut {NoStop}%
\bibitem [{\citenamefont {Collett}\ \emph {et~al.}(1987)\citenamefont
  {Collett}, \citenamefont {Loudon},\ and\ \citenamefont
  {Gardiner}}]{collett1987}%
  \BibitemOpen
  \bibfield  {author} {\bibinfo {author} {\bibfnamefont {M.~J.}\ \bibnamefont
  {Collett}}, \bibinfo {author} {\bibfnamefont {R.}~\bibnamefont {Loudon}}, \
  and\ \bibinfo {author} {\bibfnamefont {C.~W.}\ \bibnamefont {Gardiner}},\
  }\href@noop {} {\bibfield  {journal} {\bibinfo  {journal} {J. Mod. Opt.}\
  }\textbf {\bibinfo {volume} {34}},\ \bibinfo {pages} {881} (\bibinfo {year}
  {1987})}\BibitemShut {NoStop}%
\bibitem [{\citenamefont {Carmichael}(1987)}]{carmichael1987}%
  \BibitemOpen
  \bibfield  {author} {\bibinfo {author} {\bibfnamefont {H.~J.}\ \bibnamefont
  {Carmichael}},\ }\href@noop {} {\bibfield  {journal} {\bibinfo  {journal} {J.
  Opt. Soc. Am. B}\ }\textbf {\bibinfo {volume} {4}},\ \bibinfo {pages} {1588}
  (\bibinfo {year} {1987})}\BibitemShut {NoStop}%
\bibitem [{\citenamefont {Ou}\ \emph {et~al.}(1987)\citenamefont {Ou},
  \citenamefont {Hong},\ and\ \citenamefont {Mandel}}]{ou1987}%
  \BibitemOpen
  \bibfield  {author} {\bibinfo {author} {\bibfnamefont {Z.~Y.}\ \bibnamefont
  {Ou}}, \bibinfo {author} {\bibfnamefont {C.~K.}\ \bibnamefont {Hong}}, \ and\
  \bibinfo {author} {\bibfnamefont {L.}~\bibnamefont {Mandel}},\ }\href@noop {}
  {\bibfield  {journal} {\bibinfo  {journal} {J. Opt. Soc. Am. B}\ }\textbf
  {\bibinfo {volume} {4}},\ \bibinfo {pages} {1574} (\bibinfo {year}
  {1987})}\BibitemShut {NoStop}%
\bibitem [{\citenamefont {Caves}\ and\ \citenamefont
  {Drummond}(1994)}]{caves1994}%
  \BibitemOpen
  \bibfield  {author} {\bibinfo {author} {\bibfnamefont {C.~M.}\ \bibnamefont
  {Caves}}\ and\ \bibinfo {author} {\bibfnamefont {P.~D.}\ \bibnamefont
  {Drummond}},\ }\href@noop {} {\bibfield  {journal} {\bibinfo  {journal} {Rev.
  Mod. Phys.}\ }\textbf {\bibinfo {volume} {66}},\ \bibinfo {pages} {481}
  (\bibinfo {year} {1994})}\BibitemShut {NoStop}%
\bibitem [{\citenamefont {Ou}\ and\ \citenamefont {Kimble}(1995)}]{ou1995}%
  \BibitemOpen
  \bibfield  {author} {\bibinfo {author} {\bibfnamefont {Z.~Y.}\ \bibnamefont
  {Ou}}\ and\ \bibinfo {author} {\bibfnamefont {H.~J.}\ \bibnamefont
  {Kimble}},\ }\href@noop {} {\bibfield  {journal} {\bibinfo  {journal} {Phys.
  Rev. A}\ }\textbf {\bibinfo {volume} {52}},\ \bibinfo {pages} {3126}
  (\bibinfo {year} {1995})}\BibitemShut {NoStop}%
\bibitem [{\citenamefont {Haus}(2011)}]{haus2011}%
  \BibitemOpen
  \bibfield  {author} {\bibinfo {author} {\bibfnamefont {H.~A.}\ \bibnamefont
  {Haus}},\ }\href@noop {} {\emph {\bibinfo {title} {Electromagnetic Noise and
  Quantum Optical Measurements}}}\ (\bibinfo  {publisher} {Springer-Verlag
  Berlin Heidelberg},\ \bibinfo {year} {2011})\BibitemShut {NoStop}%
\bibitem [{\citenamefont {Caves}(1982)}]{caves1982}%
  \BibitemOpen
  \bibfield  {author} {\bibinfo {author} {\bibfnamefont {C.~M.}\ \bibnamefont
  {Caves}},\ }\href@noop {} {\bibfield  {journal} {\bibinfo  {journal} {Phys.
  Rev. D}\ }\textbf {\bibinfo {volume} {26}},\ \bibinfo {pages} {1817}
  (\bibinfo {year} {1982})}\BibitemShut {NoStop}%
\bibitem [{\citenamefont {A.~M.~Marino}\ \emph {et~al.}(2007)\citenamefont
  {A.~M.~Marino}, \citenamefont {Stroud}, \citenamefont {Wong}, \citenamefont
  {Bennink},\ and\ \citenamefont {Boyd}}]{marino2007}%
  \BibitemOpen
  \bibfield  {author} {\bibinfo {author} {\bibfnamefont {A.}~\bibnamefont
  {A.~M.~Marino}}, \bibinfo {author} {\bibfnamefont {C.~R.}\ \bibnamefont
  {Stroud}}, \bibinfo {author} {\bibfnamefont {V.}~\bibnamefont {Wong}},
  \bibinfo {author} {\bibfnamefont {R.~S.}\ \bibnamefont {Bennink}}, \ and\
  \bibinfo {author} {\bibfnamefont {R.~W.}\ \bibnamefont {Boyd}},\ }\href@noop
  {} {\bibfield  {journal} {\bibinfo  {journal} {J. Opt. Soc. Am. B}\ }\textbf
  {\bibinfo {volume} {24}},\ \bibinfo {pages} {335} (\bibinfo {year}
  {2007})}\BibitemShut {NoStop}%
\bibitem [{\citenamefont {Glauber}(1963{\natexlab{a}})}]{glauber1963b}%
  \BibitemOpen
  \bibfield  {author} {\bibinfo {author} {\bibfnamefont {R.~J.}\ \bibnamefont
  {Glauber}},\ }\href@noop {} {\bibfield  {journal} {\bibinfo  {journal} {Phys.
  Rev.}\ }\textbf {\bibinfo {volume} {131}},\ \bibinfo {pages} {2766} (\bibinfo
  {year} {1963}{\natexlab{a}})}\BibitemShut {NoStop}%
\bibitem [{\citenamefont {Mandel}\ and\ \citenamefont
  {Wolf}(1995)}]{mandel1995}%
  \BibitemOpen
  \bibfield  {author} {\bibinfo {author} {\bibfnamefont {L.}~\bibnamefont
  {Mandel}}\ and\ \bibinfo {author} {\bibfnamefont {E.}~\bibnamefont {Wolf}},\
  }\href@noop {} {\emph {\bibinfo {title} {Optical Coherence and Quantum
  Optics}}},\ \bibinfo {edition} {1st}\ ed.,\ Vol.~\bibinfo {volume} {1}\
  (\bibinfo  {publisher} {Cambridge University Press},\ \bibinfo {address} {New
  York city, New York},\ \bibinfo {year} {\noopsort{1973c}1995})\BibitemShut
  {NoStop}%
\bibitem [{\citenamefont {Fan}\ \emph {et~al.}(2014)\citenamefont {Fan},
  \citenamefont {He},\ and\ \citenamefont {Feng}}]{fan2014}%
  \BibitemOpen
  \bibfield  {author} {\bibinfo {author} {\bibfnamefont {H.}~\bibnamefont
  {Fan}}, \bibinfo {author} {\bibfnamefont {D.}~\bibnamefont {He}}, \ and\
  \bibinfo {author} {\bibfnamefont {S.}~\bibnamefont {Feng}},\ }\href@noop {}
  {\enquote {\bibinfo {title} {Quantum noise in phase-sensitive heterodyne
  detection with a bichromatic local oscillator},}\ }\bibinfo {howpublished}
  {e-print arXiv:1410.8602} (\bibinfo {year} {2014})\BibitemShut {NoStop}%
\bibitem [{\citenamefont {Feng}\ \emph {et~al.}(2012)\citenamefont {Feng},
  \citenamefont {Lu}, \citenamefont {Zhang},\ and\ \citenamefont
  {Shao}}]{feng2012}%
  \BibitemOpen
  \bibfield  {author} {\bibinfo {author} {\bibfnamefont {S.}~\bibnamefont
  {Feng}}, \bibinfo {author} {\bibfnamefont {Z.~H.}\ \bibnamefont {Lu}},
  \bibinfo {author} {\bibfnamefont {J.}~\bibnamefont {Zhang}}, \ and\ \bibinfo
  {author} {\bibfnamefont {C.~G.}\ \bibnamefont {Shao}},\ }\href@noop {}
  {\enquote {\bibinfo {title} {Balanced-heterodyne detection of sub-shot-noise
  optical signals},}\ }\bibinfo {howpublished} {e-print arXiv:1112.3155v2}
  (\bibinfo {year} {2012})\BibitemShut {NoStop}%
\bibitem [{\citenamefont {Drummond}\ and\ \citenamefont
  {Reid}(1990)}]{drummond90}%
  \BibitemOpen
  \bibfield  {author} {\bibinfo {author} {\bibfnamefont {P.~D.}\ \bibnamefont
  {Drummond}}\ and\ \bibinfo {author} {\bibfnamefont {M.~D.}\ \bibnamefont
  {Reid}},\ }\href@noop {} {\bibfield  {journal} {\bibinfo  {journal} {Phys.
  Rev. A}\ }\textbf {\bibinfo {volume} {41}},\ \bibinfo {pages} {3930}
  (\bibinfo {year} {1990})}\BibitemShut {NoStop}%
\bibitem [{\citenamefont {Glauber}(1963{\natexlab{b}})}]{glauber1963a}%
  \BibitemOpen
  \bibfield  {author} {\bibinfo {author} {\bibfnamefont {R.~J.}\ \bibnamefont
  {Glauber}},\ }\href@noop {} {\bibfield  {journal} {\bibinfo  {journal} {Phys.
  Rev.}\ }\textbf {\bibinfo {volume} {130}},\ \bibinfo {pages} {2529} (\bibinfo
  {year} {1963}{\natexlab{b}})}\BibitemShut {NoStop}%
\end{thebibliography}%

%\end{CJK}
\end{document}